\documentclass[twoside,slac_one]{revtex4}
\usepackage{graphicx}
\usepackage{fancyhdr}
\usepackage{amsmath} 
\usepackage{bm}
\usepackage{amsxtra}
\usepackage{amssymb}
\usepackage{amsthm}
\usepackage{latexsym}
\usepackage{lscape}

\begin{document}

\title{Bringing the LHC and ATLAS to a regional planetarium}

%

\author{R. Schwienhorst}
\affiliation{Department of Physics and Astronomy, Michigan State University, East Lansing, USA}

\begin{abstract}
An outreach effort has started at Michigan State University to bring particle 
physics, the Large Hadron Collider, and the ATLAS experiment to a general 
audience at the Abrams planetarium on the MSU campus. 
A team of undergraduate students majoring 
in physics, communications arts \& sciences, and journalism are putting 
together short clips about ATLAS and the LHC to be shown at the planetarium.  
\end{abstract}

\maketitle

\thispagestyle{fancy}


\section{Introduction}
The two general-purpose LHC experiments ATLAS~\cite{atlasoutreach} and
CMS~\cite{CMSoutreach} have strong outreach programs that target audiences interested 
in particle physics and the LHC. An additional outreach tool not yet utilized by
the LHC experiments is the planetarium. There are thousands of planetariums world-wide 
providing education and entertainment to a wide variety of target audiences, from 
elementary school groups to university classes and general audiences. The Abrams 
planetarium~\cite{abrams} on the campus of Michigan State University for example serves
about 35,000 visitors annually, many of whom have expressed curiosity about how the
LHC will address some of the important questions in astronomy and cosmology. An outreach
program designed to bring the LHC and in particular the ATLAS experiment to the Abrams
planetarium has been started at Michigan State University. The first result of this
effort is a short planetarium clip about particle physics, the LHC and ATLAS and how
MSU contributes which is shown in the planetarium.

\section{Abrams planetarium}
\begin{figure}[ht]
\centering
\includegraphics[width=80mm]{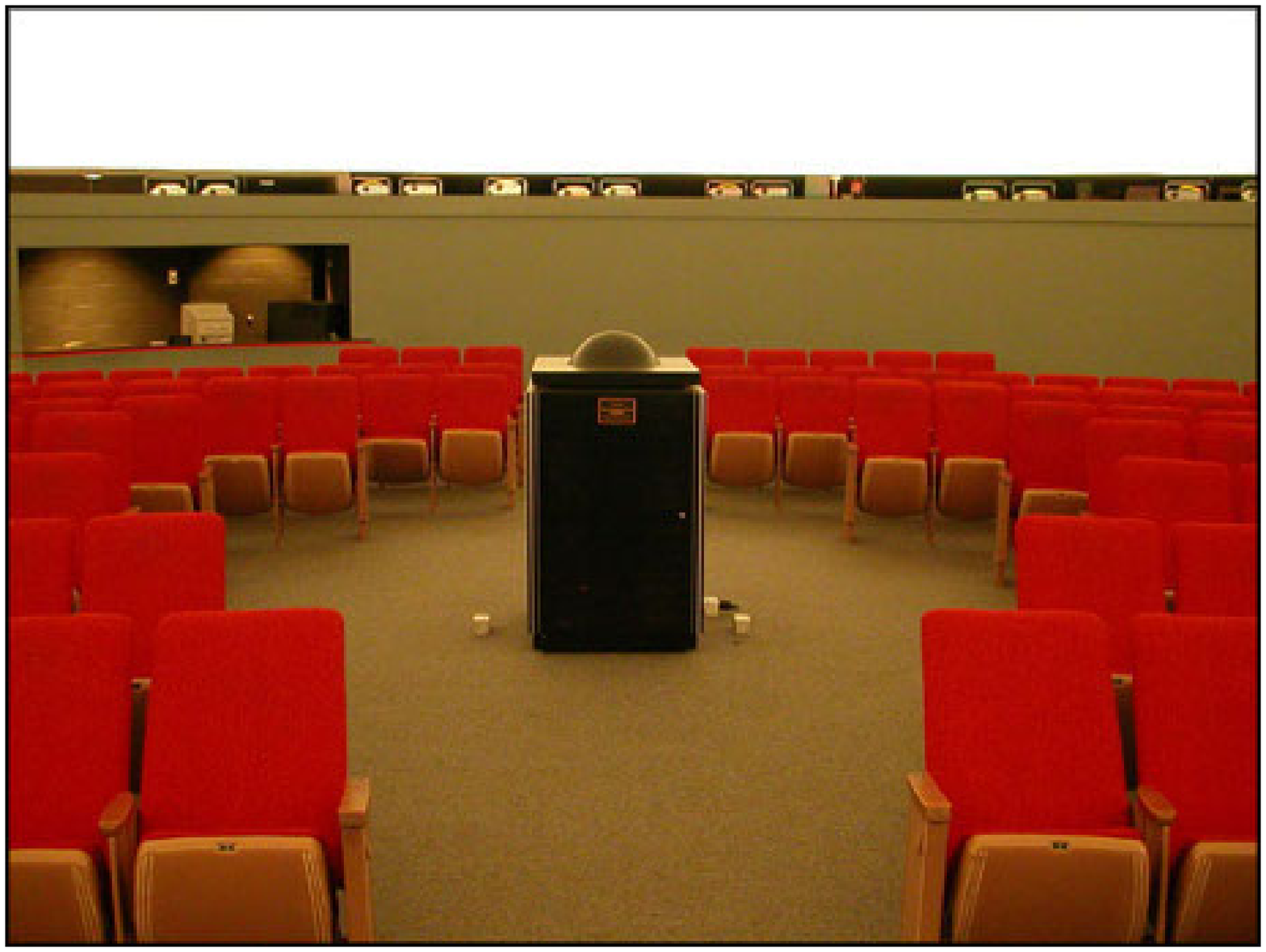}
\includegraphics[width=80mm]{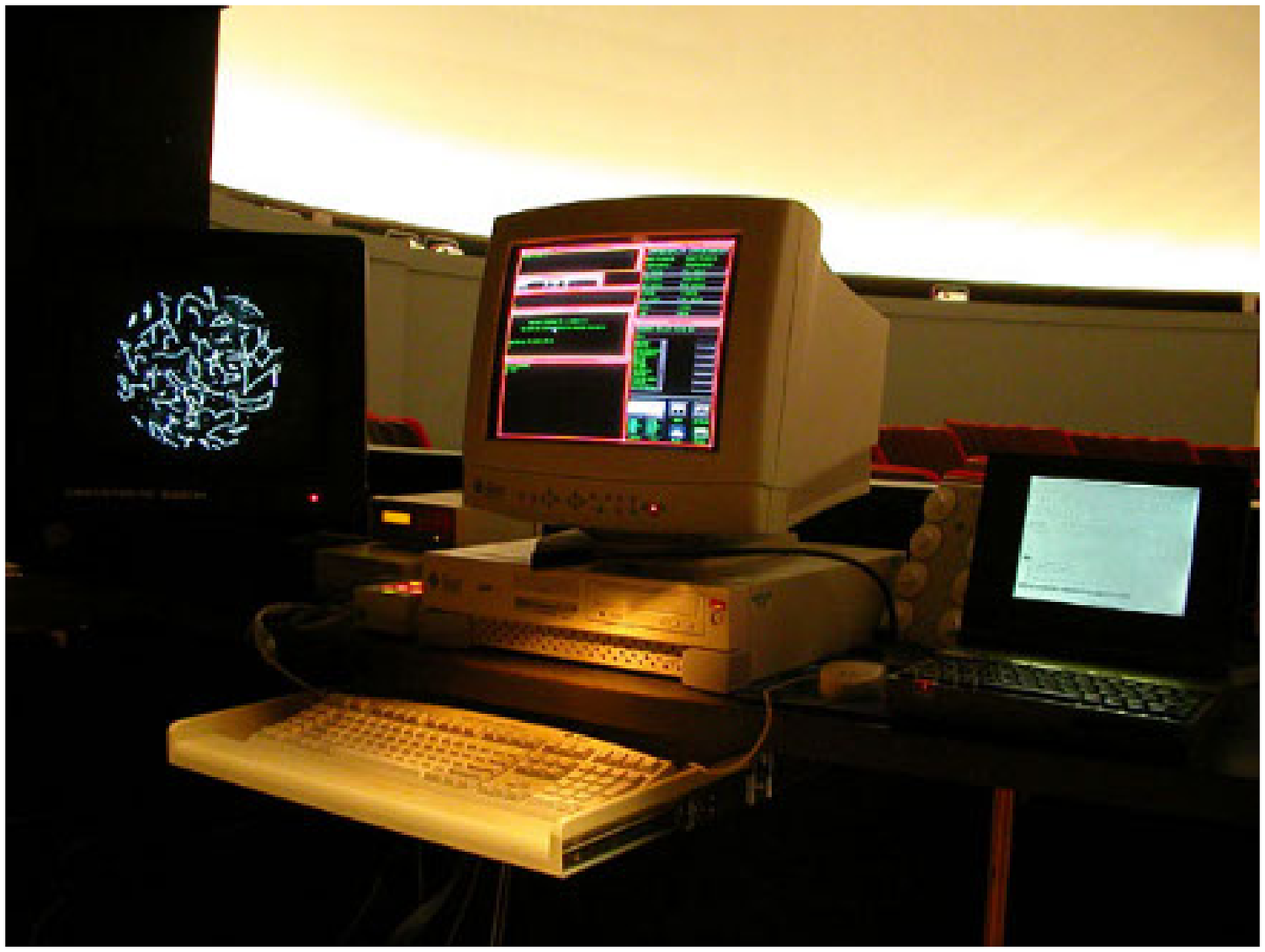}
\caption{Digistar II projector inside Abrams planetarium at Michigan State
University and its control console.} \label{fig:digistar}
\end{figure}

The Abrams planetarium is part of the Department of Physics \& Astronomy at 
Michigan State University and serves as an astronomy and space science education 
resource, providing courses for MSU students, educational shows for area schools, 
and weekend shows for the general public. About 35,000 people visit the planetarium 
annually. The planetarium uses the Evans \& Sutherland Digistar II projector 
system~\cite{digistar2}, consisting of a CRT and a lens to provide full-dome 
monochrome images. The CRT is controlled by a computer (Sun UNIX workstation) with a 
graphics processor that renders vector graphics. This computer is also connected to a 
LCD projector installed in the planetarium. The system is additionally connected to
several slide projectors that can project on various areas of the dome, including 
full-dome images, as well as a video projector and several special-effects projectors.
The Digistar~II projector at its location in the planetarium as well as its control 
station are shown in Fig.~\ref{fig:digistar}.
~

The video projector and the various slide projectors in the system are well
suited to display particle physics information, and even the Digistar~II projector
can show particle collisions and accelerator parts or detectors as line-drawings. 

\begin{figure}[ht]
\centering
\includegraphics[width=135mm,height=135mm]{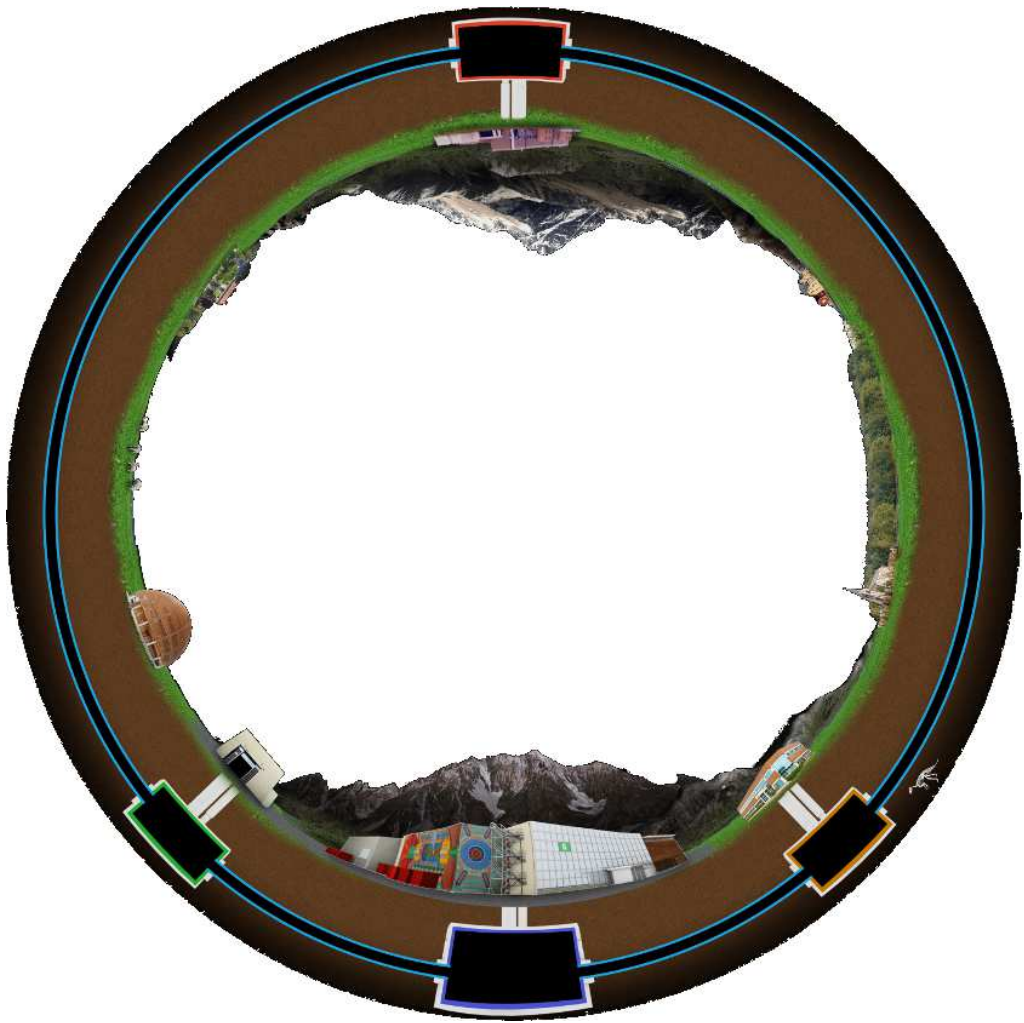}
\caption{All-sky image of the LHC accelerator and the detector sites, together with
the surrounding mountains and countryside. This image is projected onto the 
planetarium dome, with the bottom of the image appearing in the front.} 
\label{fig:allsky}
\end{figure}

\section{Student involvement}
Michigan State University has a graphics information program, a graphic design program,
and various courses teaching undergraduate students about state-of-the-art animation,
image creation, graphics processing, and audio production. Four undergraduate students 
from the College of Communication Arts \& Sciences~\cite{msucas} are working on 
creating animations and images about the ATLAS experiment, the LHC, and particle 
physics. 
Several of the students and their adviser traveled to CERN in December 2010 to see 
and experience first-hand what goes on at the LHC. The visit was very useful
to the students and facilitated the production of the short planetarium clip. 
The students brought knowledge and material (images and videos) back with them, both
of which are now being used.

At MSU, the animators and graphic designers are working together with an audio
production student, two students from the professional writing program,
as well as a physics graduate student and a physics undergrad.
This team is responsible for creating planetarium show modules, from putting 
the initial script 
together to creating the animations and images and assembling the pieces into a 
coherent self-contained piece. The team works well together despite the very different
backgrounds of the individuals involved. 
This creative experience benefits all of the students. The main benefit to the
physics students is much improved communication skills, while the main benefit 
for the graphic designers and animators is a better understanding of how science
works and how to communicate it.

The students use comercial graphic design and animation software packages on 
dedicated high-end graphics workstations. The students are well-trained in the 
required software through their course-work and other projects. This includes 3D 
modeling and animation software, video editing and production programs, graphic 
design and image processing packages, and audio production programs.

While working mostly independently, the undergraduate students receive guidance
from professors in the Information Graphics and in the Professional Writing programs at 
MSU. This connection, in addition to the support from Abrams planetarium staff, is
useful to ensure that all aspects of the planetarium clip design process are
covered and that the final result will be comprehensible to a general planetarium 
audience.

\begin{figure}[ht]
\centering
\includegraphics[width=135mm,height=135mm]{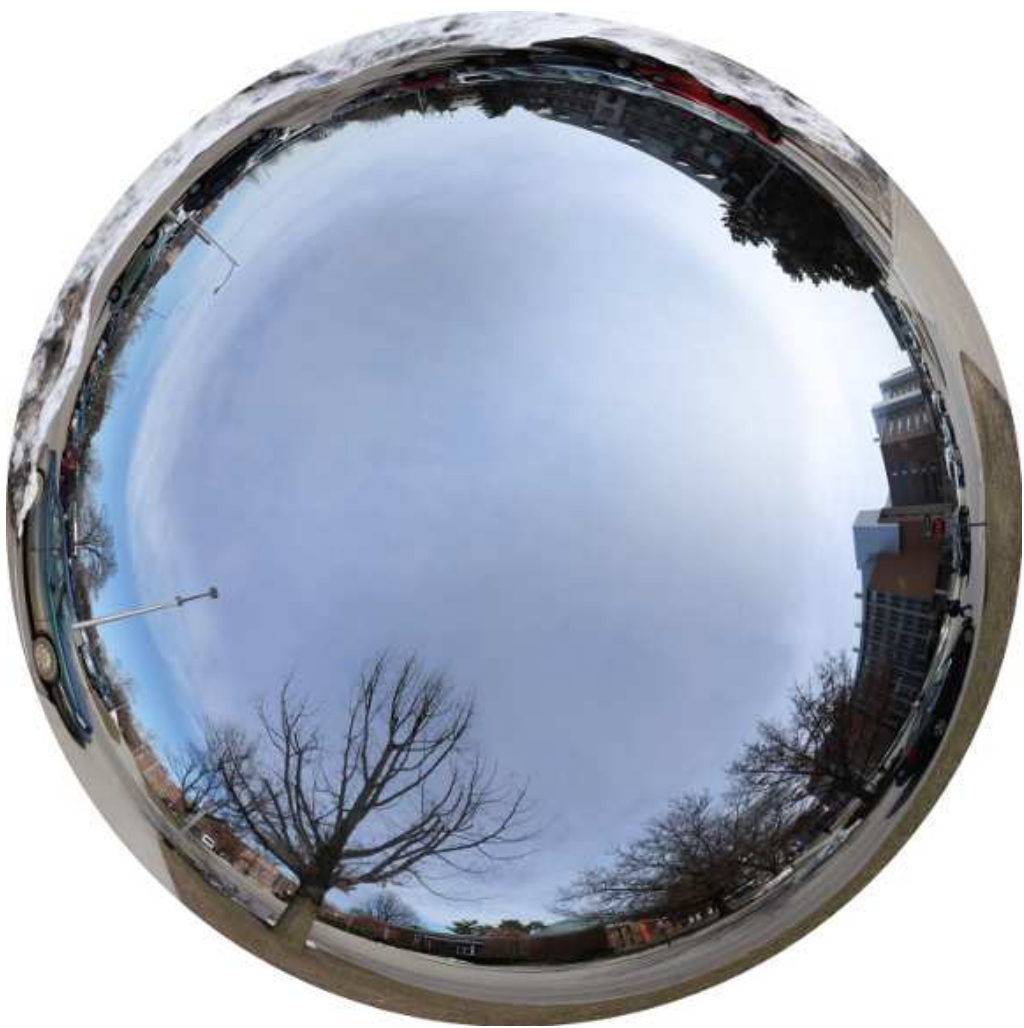}
\caption{All-sky image of the Michigan State University campus, showing Abrams
planetarium at the bottom (in the front of the dome) and the Biomedical and Physical
Sciences building on the right.} 
\label{fig:bps}
\end{figure}

\section{Short clip}
The creative team at MSU has spent the spring semester 2011 on creating a first 
five-minute long clip~\cite{shortclip}. This clip plays at the end of each public
Abrams planetarium show. The current public show is on the topic of dark 
matter, and the five-minute clip describes how the LHC and ATLAS expect to identify
dark matter and how MSU students and researchers contribute to this effort. 
An example all-sky image from the short clips is shown in 
Fig.~\ref{fig:allsky}. All-sky images such as this one are projected onto the dome
such that the bottom part of the image appears in the front, the top of the image
appears at the back, and the center of the image is directly overhead. This particular
image is combined with a Digistar animation of counter-circulating beams inside the
LHC tunnel. The Digistar projector also shows stars in the night sky overhead.

Another example all-sky image is shown in Fig.~\ref{fig:bps}. This image of the
MSU campus with the planetarium and the physics building makes it easy for visitors
to orient themselves and get a sense of where the researchers work.

Several other images and animations have been created in addition to the images
shown in Figs.~\ref{fig:allsky} and~\ref{fig:bps}. All content is created locally
at MSU by the students. This process not only results in content that the creative
team and the audience can identify with, but it also builds a repository of 
MSU-specific content that can also be used elsewhere. It also gives future students 
the option and flexibility to build on previous work.

\section{Feedback}
After each showing of the short clip, the physicist members of the creative team are 
available to the audience for questions. The feedback on the short clip collected in 
this way has been overwhelmingly positive and encouraging. Many planetarium visitors 
are interested in astronomy, astrophysics and cosmology. While they are aware of the
existence of the LHC, they lack an understanding of specific questions that the LHC 
will address and how it will address them. The short clip is well suited to address
these questions. 

Additionally, feedback from MSU students and alumni on the connection between
MSU and the LHC has also been very positive. Several people expressed surprise at how 
much individual students and researchers from a university like MSU are able to 
contribute to such a large and complex project.

\section{Touch screen}
\begin{figure}[ht]
\centering
\includegraphics[width=135mm]{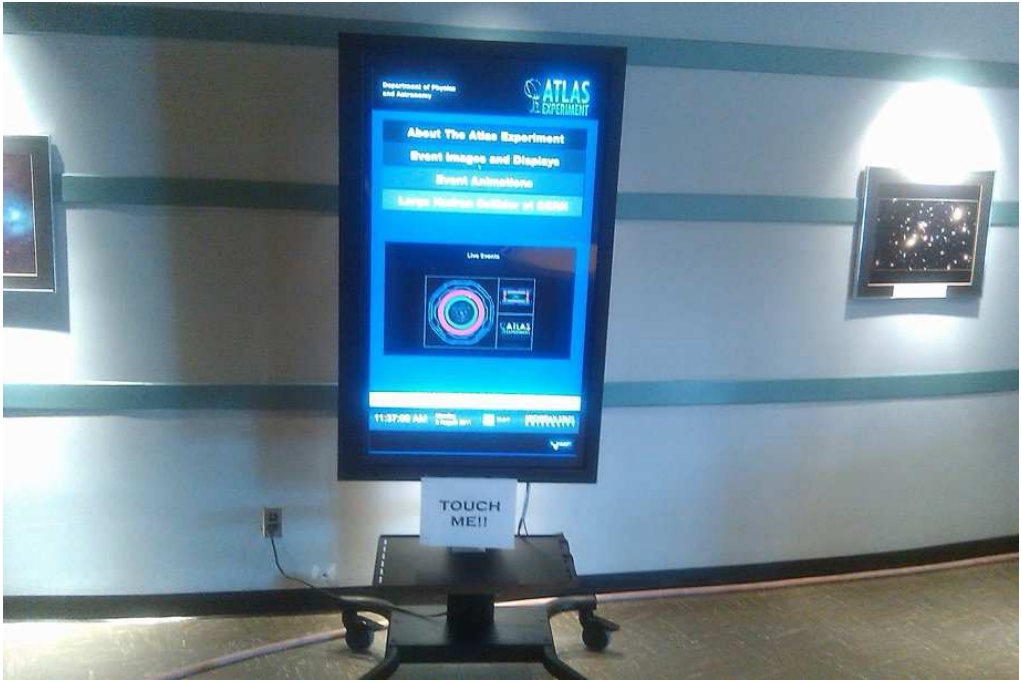}
\caption{Dynics touch screen in the Abrams planetarium lobby.} \label{fig:touchscreen}
\end{figure}

The images and animations produced by the creative team can not only be shown inside
the planetarium dome but will also be useful for other outreach projects. For example,
a Dynics touch screen has been placed in the planetarium lobby with the intention to
display particle physics and ATLAS outreach content~\ref{fig:touchscreen}. The display
contains images, videos and background articles on ATLAS and the LHC. It also shows the
live ATLAS event display web page~\cite{atlaslive}.
By converting the images, animations and other content developed for the planetarium 
into a format suitable for the touch screen we can deepen the viewers experience
and reinforce the educational content. Curious visitors get an opportunity to 
learn more about ATLAS and the LHC and to explore on their own. It is typically the
younger audience who members feel more comfortable to approach the touch screen and 
start exploring it, pushing the various on-screen buttons and interacting with the
display. Combining general ATLAS-wide outreach material with MSU-specific content
developed for the planetarium show makes the touch screen display an excellent
outreach tool.
~

\section{Conclusions}
A team of undergraduate students at Michigan State University has produced a short
clip on particle physics, the LHC and the ATLAS experiment for the Abrams 
planetarium on the MSU campus. This five-minute clip has been a big success with
planetarium audiences.
The creative team is now working on a full planetarium show that
will premiere as the feature show at the Abrams planetarium in 
Fall 2011~\cite{relics}. This show will be about 35 minutes long and cover the
connection between the LHC and the ATLAS experiment on one side and cosmology and the
big bang on the other, with a special emphasis on MSU's involvement.
It will feature the same mix of all-sky and slide images, animations and digistar
programming also present in the short clip. 
Show modules and clips will also be made available
to other planetariums that have a classic setup of slide and video projectors similar
to Abrams planetarium.

\begin{acknowledgments}
This work is supported in part by NSF grants PHY-0952729 and PHY-0757741.
\end{acknowledgments}

\bigskip 

\end{document}